\pdfoutput=1
\documentclass[prb,twocolumn,showpacs,amsmath,amssymb,floatfix,superscriptaddress]{revtex4}

\usepackage{graphicx}%Include figure files
\usepackage{dcolumn}%Align table columns on decimal point
\usepackage{bm}% bold math
\usepackage{amsmath}
\usepackage{amssymb}
\newcommand*{\Tr}{%
	\mathrm{Tr} }

\begin{document}

	\title{Intrinsic Damping Phenomena from Quantum to Classical Magnets: \\
		An ab-initio Study of Gilbert Damping in Pt/Co Bilayer}
	
	\author{Farzad Mahfouzi}
	\email{Farzad.Mahfouzi@gmail.com}
	\affiliation{Department of Physics and Astronomy, California State University, Northridge, CA, USA}
	\author{Jinwoong Kim}
	\affiliation{Department of Physics and Astronomy, California State University, Northridge, CA, USA}
	\affiliation{Department of Physics and Astronomy, Rutgers University, NJ, USA}
	\author{Nicholas Kioussis}
	\email{nick.Kioussis@csun.edu }
	\affiliation{Department of Physics and Astronomy, California State University, Northridge, CA, USA}
	
	\begin{abstract}
		
	A fully quantum mechanical description of the precessional damping of Pt/Co bilayer is presented in the framework of the Keldysh Green function approach using {\it ab initio} electronic structure calculations. In contrast to previous calculations of classical Gilbert damping ($\alpha_{GD}$), we demonstrate that $\alpha_{GD}$ in the quantum case does not diverge in the ballistic regime due to the finite size of the total spin, $S$.  In the limit of $S\rightarrow\infty$ we show that the formalism recovers the torque correlation expression for $\alpha_{GD}$  which we decompose into spin-pumping and spin-orbital torque correlation contributions. The formalism is generalized to take into account a self consistently determined dephasing mechanism which preserves the conservation laws and allows the investigation of the effect of disorder. The dependence of $\alpha_{GD}$ on Pt thickness and disorder strength is calculated and the spin diffusion length of Pt and spin mixing conductance of the bilayer are determined and compared with experiments.
	\end{abstract}

	\pacs{72.25.Mk, 75.70.Tj, 85.75.-d, 72.10.Bg}
	\maketitle
	
	\section{Introduction}\label{sec:intro}
	
	Magnetic materials provide an intellectually rich arena for fundamental scientific discovery and
	for the invention of faster, smaller and more energy-efficient technologies. The intimate relationship of charge
	transport and magnetic structure in metallic systems on one hand, and the rich physics occurring at the interface between different materials in layered structures on the other hand, are the hallmark of the
	flourishing research field of spintronics. \cite{Miron2010,Miron2011,Liu2012,Liu2012_1,Zutic2004}

Recently, intense focus has been placed on the significant role played by spin-orbit coupling (SOC) and the effect of  interfacial inversion symmetry breaking on the dynamics of the magnetization in ferromagnet (FM)-normal metal (NM) bilayer systems. Of prime importance to this field
is the (precessional) magnetization damping phenomena, usually	treated phenomenologically by means of a parameter referred to as Gilbert damping constant, $\alpha_{GD}$, in the Landau–Lifshitz–Gilbert  (LLG) equation of motion $d\vec{m}/dt=\gamma\vec{m}\times \vec{B}+\alpha_{GD}\vec{m}\times d\vec{m}/dt$, which describes the rate of the angular momentum loss of the FM.\cite{Gilbert2004}
	Here, $\vec{m}$ is the unit vector along the magnetization direction and $\vec{B}$ is an effective magnetic field.
	
	In FM/NM bilayer devices the effect of the NM on the Gilbert damping of the FM is typically considered as an additive effect, where the total Gilbert damping can be separated into an intrinsic bulk contribution and an interfacial component due to the presence of the NM.\cite{Berger1996,Barati2014} While the interfacial Gilbert damping is usually attributed to the loss of angular momentum due to pumped spin current into the NM,\cite{Tserkovnyak2002,Zwierzycki2005} in metallic bulk FMs the intrinsic Gilbert damping constant is described by the coupling between the conduction electrons and the (time-dependent) magnetization degree of freedom.\cite{Kambersky2007}
	
	The conventional approach to determine the Gilbert damping constant involves calculating the imaginary part of the time-dependent susceptibility of the FM in the presence of conduction electrons in the linear response regime. \cite{Garate2009_1,Mills2003,Simanek2003} In this case, the time-dependent magnetization term in the electronic Hamiltonian leads to the excitation of electrons close to the Fermi surface transferring angular momentum to the conduction electrons. The excited electrons in turn relax to the ground state by interacting with their environment, namely through phonons, photons and/or collective spin/charge excitations. These interactions are typically parameterized phenomenologically by the broadening of the energy levels, $\eta=\hbar/2\tau$, where $\tau$ is the relaxation time of the electrons close to the Fermi surface. The phenomenological treatment of the electronic relaxation is  valid when the energy broadening is small which corresponds to clean systems,
	i.e., $\eta D(E_F)\leq 1$, where $D(E_F)$ is the density of states per atom at the Fermi energy.
	In the case of large $\eta$ [$\eta D(E_F)\gtrsim 1$)] however, this approach violates the conservation laws and a more accurate description of the relaxation mechanism that preserves the energy, charge and angular momentum conservation laws are required.\cite{GBaym1961} The importance of including the vertex corrections  has already been pointed out in the literature when the Gilbert damping is dominated by the interband contribution,\cite{Garate2009,Mankovsky2013,Turek2015} {\it i.e.,} when there is a significant number of states available within the energy window of $\eta$ around the Fermi energy.

%%%%%%%%%%%%%%%%%%%%%%%%%%%%%%%%%%%%%%%%%%%%%%%%%%%%  OBJECTIVE %%%%%%%%%%%%%%%%%%%%%%%%%%%%%%%%%%%%%%%%%	
	In this paper we investigate the magnetic damping phenomena through a different Lens in which the FM is assumed to be small and quantum mechanical. We show that in the limit of large magnetic moments we recover different conventional expressions for the Gilbert damping of a classical FM. We calculate the Gilbert damping for a Pt/Co bilayer system versus the energy broadening, $\eta$ and show that in the limit of clean systems and small magnetic moments the FM damping is governed by a coherent dynamics. We show that in the limit of large broadening $\eta>1 meV$ which is typically the case at room temperature, the relaxation time approximation fails. Hence, we employ a self consistent approach preserving the conservation laws. We calculate the Gilbert damping versus the Pt and Co thicknesses and by fitting the results to spin diffusion model we calculate the spin diffusion length and spin mixing conductance of Pt.

%%%%%%%%%%%%%%%%%%%%%%%%%%%%%%%%%%%%%%%%%%%%%%%%%%%%%%%%%%%%%%%%%%%%%%%%%%%%%%%%%%%%%%%%%
	\section{Theoretical Formalism of Magnetization Damping}
	For a metallic FM the magnetization degree of freedom is inherently coupled to the electronic degrees of freedom of the conduction electrons. It is usually convenient to treat each degree of freedom separately with the corresponding time-dependent Hamiltonians that do not conserve the energy. However, since the total energy of the system is conserved, it is  possible to consider the total Hamiltonian of the combined system and solve the corresponding stationary equations of motion. For an isolated metallic FM the wave function of the coupled electron-magnetic moment configuration system is of the form, $|m\alpha\vec{k}\rangle=|S,m\rangle\otimes |\alpha\vec{k}\rangle$, where the parameter $S$ denotes the total spin of the nano-FM ($S\rightarrow\infty$ in the classical limit),
	$m = -S \ldots ,+S$, are the eigenvalues of the total S$_z$ of
	the nano-FM, $\otimes$ refers to the Kronecker product, and $\alpha$ denotes the atomic orbitals and spin of the electron Bloch states. The single-quasi-particle retarded Green function and the corresponding density matrix can be obtained from,\cite{Mahfouzi2017}
	\begin{align}\label{eq:GFeq1}
	&\left(E-i\eta-\hat{H}_{\vec{k}}-\boldsymbol{H}_M-\frac{1}{2S}\hat{\Delta}_{\vec{k}}\hat{\vec{\sigma}}\cdot\vec{\boldsymbol{S}}\right)\hat{\boldsymbol{G}}_{\vec{k}}^r(E)=\hat{\boldsymbol{1}},
	\end{align}
	and
	\begin{align}\label{eq:GFeq2}
	&\hat{\boldsymbol{\rho}}_{\vec{k}}=\int \frac{dE}{\pi}\hat{\boldsymbol{G}}_{\vec{k}}^r(E)\eta {f}(E-\boldsymbol{H}_M)\hat{\boldsymbol{G}}_{\vec{k}}^a(E).
	\end{align}
	Here, $\boldsymbol{H}_M=\gamma\vec{B}\cdot\vec{\boldsymbol{S}}$, is the Hamiltonian of
	the nano-FM in the presence of an external magnetic field $\vec{B}$ with eigenstates, $|S,m\rangle$,
	 $\gamma$ is the gyromagnetic ratio, $f(E)$ is the Fermi-Dirac distribution function,
	 $\hat{\vec{\sigma}}$ is the vector of the Pauli matrices,
	 $\hat{H}_{\vec{k}}$ is the non-spin-polarized Hamiltonian matrix
	 in the presence of spin orbit coupling (SOC), and $\hat{\Delta}_{\vec{k}}$ is the $\vec{k}$-dependent
	 exchange splitting matrix, discussed in detail in Sec. III.
	 We employ the notation that bold symbols operate on $|S,m\rangle$ basis set and symbols with hat operate on the $|\alpha\vec{k}\rangle$s. Here, for simplicity we ignore explicitly writing the identity matrices $\hat{1}$ and $\bold{1}$ as well as the Kronecker product symbol in the expressions.
	
%%%%%%%%%%%%%%%%%%%%%%%%%%%%%%%%%

\begin{figure}
		\includegraphics[scale=0.4,angle=0]{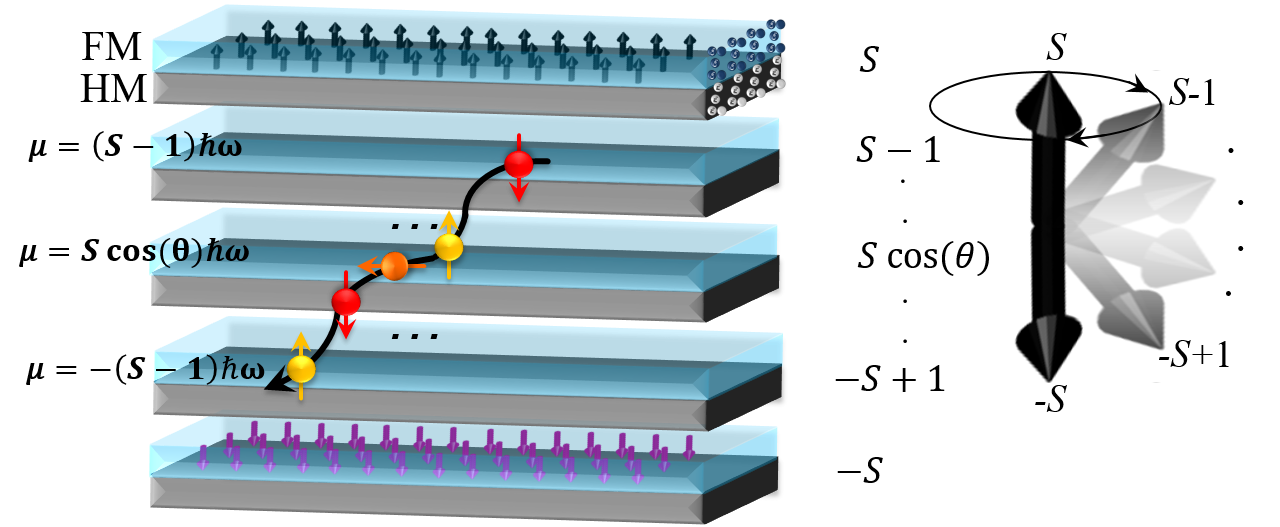}
		\caption{(Color online) Schematic representation of the combined FM-Bloch electron system. The horizontal planes denote the eigenstates, $|S,m\rangle$
			of the total $\boldsymbol{S}_z$ of the nano-FM with eigenvalues $m = -S, -S+1,\ldots ,+S$. For more details see Fig. 2 in Ref. \cite{Mahfouzi2017}}.
		\label{fig:fig1}
	\end{figure}

  A schematic description of the FM-Bloch electron entangled system and the damping process of the nano-FM  is shown in Fig. \ref{fig:fig1}. The presence of the magnetic Hamiltonian in the Fermi distribution function in Eq. (\ref{eq:GFeq2}) acting as a chemical potential leads to transition between magnetic states
  $|S,m\rangle$  along the direction in which the magnetic energy is minimized\cite{Mahfouzi2017}. The transition rate of the FM from the excited states, $|S,m\rangle$, to states with lower energy ({\it i.e.} the damping rate) can be calculated from\cite{Mahfouzi2017},
%%%%%%%%%%%%%%%%%%%%%%%%%%%%%%%%%%%%%%%%%%%
	\begin{align}
	\mathcal{T}_m&=\frac{1}{2}\Im(\mathcal{T}^-_{m}-\mathcal{T}^+_{m}),
	\label{eq:DampRate}
	\end{align}
	where,
	\begin{align}
	\mathcal{T}^{\pm}_{m}&=\frac{1}{2S\mathcal{N}}\sum_{\vec{k}}Tr_{el}[ \hat{\Delta}_{\vec{k}}\hat{\sigma}^{\mp}{\boldsymbol{S}}^{\pm}_m\hat{\boldsymbol{\rho}}_{\vec{k};m,m\pm 1}].
	\end{align}
	Here, $\mathcal{N}$ is the number of $\vec{k}$-points in the first Brillouin zone, $Tr_{el}$, is the trace over the Bloch electron degrees of freedom, ${\boldsymbol{S}}_m^{\pm}=\sqrt{S(S+1)-m(m\pm1)}$, and
	$\hat{\sigma}^{\mp} \equiv \hat{\sigma}_x \mp i\hat{\sigma}_y$.
	
	The precessional Gilbert damping constant can be determined from conservation of the total angular momentum by equating the change of angular momentum per unit cell for the Bloch electrons, $\mathcal{T}_m$, and the magnetic moment obtained from LLG equation, $\alpha_{GD} M_{tot}\sin^2(\theta)/2$, which leads to,
	\begin{align}\label{eq:GD1}
	\alpha_{GD}(m)&=-\frac{2}{M_{tot}\omega\sin^2(\theta_m)}\mathcal{T}_m\nonumber\\
	&\equiv-\frac{S^2}{M_{tot}\omega(S(S+1)-m^2)}\mathcal{T}_m.
	\end{align}
	Here,  cos($\theta_m)=\frac{m}{\sqrt{S(S+1)}}$,	is the cone angle of precession	and  $M_{tot}$ is the total magnetic moment per unit cell in units of $\frac{1}{2}g\mu_B$ with $g$ and $\mu_B$ being the Land\'{e} factor and magneton Bohr respectively.  The Larmor frequency, $\omega$, can be obtained from the effective magnetic field along the precession axis, $\hbar\omega=\gamma B_z$. 
	
	The exact treatment of the magnetic degree of freedom within the single domain dynamical regime offers a more accurate description of the damping phenomena that can be used even when the classical equation of motion LLG is not applicable. 
	However, since in most cases of interest the FM behaves as a classical magnetic moment, where the adiabatic approximation can be employed to describe the magnetization dynamics, in the following two sections we consider the $S\rightarrow\infty$ limit and close to adiabatic regime for the FM dynamics.
	
	\subsection{Classical Regime: Relaxation Time Approximation}\label{sec:2B}
	
	The  dissipative component of the nonequilibrium electronic density matrix, to lowest order in $\partial/\partial t$,
	can be determined by expanding the Fermi-Dirac distribution in Eq. (\ref{eq:GFeq2}) to lowest order in
	$[\boldsymbol{H}_M]_{mm'}=\delta_{mm'}m\hbar\omega$.
	Performing a Fourier transformation with respect to the discrete Larmor frequency modes, $m\omega\equiv i\partial/\partial t$,
	we find that, $\hat{\rho}^{dis}_{neq}(t)=\frac{1}{\pi}\hbar\eta\hat{G}^ri\partial \hat{G}^a/\partial t$, where $\hat{G}^r=\big[E_F-i\eta-\hat{H}(t)\big]^{-1}$ and $\hat{G}^a=(\hat{G}^r)^{\dagger}$ are the retarded and advanced Green functions calculated at the Fermi energy, $E_F$, and a fixed time $t$.
	
	The energy absorption rate of the electrons can be determined from the expectation value of the time derivative of the electronic Hamiltonian, $E'_e=\Re(\Tr(\hat{\rho}^{dis}_{neq}(t)\partial \hat{H}/\partial t))$, where $\Re()$ refers to the real part.
	%Here we focus on energy absorption rate instead of angular momentum to keep the formalism general enough for systems other than FMs.
	Calculating the time-derivative of the Green function and using the identity, $\eta\hat{G}^r\hat{G}^a=\eta\hat{G}^a\hat{G}^r=\Im(\hat{G}^r)$, where, $\Im()$ refers to the anti-Hermitian part of the matrix, the {\it torque correlation} (TC) expression for the energy excitation rate of the electrons is of the form,
	\begin{align}\label{eq:1}
	E'_e&=\frac{\hbar}{\pi \mathcal{N}}\sum_{k}\Tr\Big[\Im(\hat{G}^r)\frac{\partial \hat{H}}{\partial t}\Im(\hat{G}^r)\frac{\partial \hat{H}}{\partial t}\Big].
	\end{align}	
	In the case of semi-infinite NM leads attached to the FM, using,  $\Im(\hat{G}^r)=\hat{G}^r\hat{\Gamma}\hat{G}^a=\hat{G}^a\hat{\Gamma}\hat{G}^r$, Eq.(\ref{eq:1}) can
	be written as
	\begin{align}\label{eq:2}
	E'_e&=\frac{\hbar}{\pi \mathcal{N}}\sum_{k}\Tr\Big[\hat{\Gamma}\frac{\partial \hat{G}^r}{\partial t}\hat{\Gamma}\frac{\partial \hat{G}^a}{\partial t}\Big]
	\end{align}	
	where, $\hat{\Gamma}=\eta\hat{1}+(\hat{\Sigma}^r-\hat{\Sigma}^a)/2i$, with $\hat{\Sigma}^{r/a}$ being the retarded$/$advanced self energy due to the NM lead attached to the FM which describes the escape rate of electrons from/to the reservoir.  It is useful to separate the dissipation phenomena into {\it local} and {\it nonlocal} components as follows. Applying the unitary operator, $\hat{U}(t)=e^{i\omega\hat{\sigma}_z t/2}e^{i\theta\hat{\sigma}_x/2}e^{-i\omega\hat{\sigma}_z t/2}=\cos(\frac{\theta}{2})\hat{1}+i\sin(\frac{\theta}{2})(\hat{\sigma}^+e^{i\omega t}+\hat{\sigma}^-e^{-i\omega t})$, to fix the magnetization orientation along $z$ we find,
		\begin{align}\label{eq:3_0}
	\frac{\partial(\hat{U}\hat{G}^r_0\hat{U}^{\dagger})}{\partial t}&\approx\frac{\omega}{2}\sin(\theta)\left(\hat{\mathcal{G}}'e^{i\omega t}+\hat{\mathcal{G}}'^{\dagger}e^{-i\omega t}\right),
	\end{align}	
	where we have ignored higher order terms in $\theta$ and,
	\begin{align}\label{eq:3_1}
	\hat{\mathcal{G}}'=[\hat{G}_0^r,\hat{\sigma}^+]-\hat{G}_0^r[\hat{H}_0,\hat{\sigma}^+]\hat{G}_0^r.
	\end{align}	
	Here, $[,]$ refers to the commutation relation, $\hat{H}_{0}$ is the time independent terms of the Hamiltonian, and $\hat{G}^{r/a}_0$ refers to the Green function corresponding to magnetization along $z$-axis. Using Eq. \eqref{eq:2} for the average energy absorption rate we obtain,
	\begin{align}\label{eq:3}
	E'_e=\frac{\hbar\omega^2}{2\pi \mathcal{N}}&\sin^2(\theta)\sum_{k}\Tr\left(\hat{\Gamma}\hat{\mathcal{G}}'\hat{\Gamma}\hat{\mathcal{G}}'^{\dagger}\right)\nonumber\\
	=-\frac{\hbar\omega^2}{2\pi \mathcal{N}}&\sin^2(\theta)\sum_{k}\Re\left(\Tr\left(\hat{\Gamma}[\hat{G}_0^r,\hat{\sigma}^+]\hat{\Gamma}[\hat{G}_0^a,\hat{\sigma}^-]\right.\right.\nonumber\\
	&+\Im(\hat{G}^r)[\hat{H}_0,\hat{\sigma}^+]\Im(\hat{G}^r)[\hat{H}_0,\hat{\sigma}^-]\nonumber\\
	&-2\left.\left.[\Im(\hat{G}_0^r),\hat{\sigma}^+]\hat{\Gamma}\hat{G}_0^a[\hat{H}_0,\hat{\sigma}^-]\right)\right).
	\end{align}	
	
	In the absence of the SOC, the first term in Eq. (\ref{eq:3}) is the only non-vanishing term which corresponds to the pumped spin current into the reservoir [i.e. $I_{S_z}=\hbar\Tr(\hat{\sigma}_z\hat{\Gamma}\hat{\rho}_{neq}^{dis})/2$] dissipated in the NM (no back flow). This spin pumping component is conventionally formulated in terms of the {\it spin mixing conductance}\cite{Tserkovnyak2005}, $I_{S_z}=\hbar g_{\uparrow\downarrow}\sin^2(\theta)/4\pi$, which
	acts as a {\it nonlocal} dissipation mechanism.
	The second term, referred to as the {\it spin-orbital torque correlation}\cite{Kamberský1970,Kambersky2007} (SOTC) expression for damping, is commonly used to calculate the intrinsic contribution to the Gilbert damping constant for bulk metallic FMs. 
	The third term arises when both SOC and the reservoir are present. It is important to note that the formalism
	presented above is valid only in the limit of small $\eta$ (ballistic regime). On the other hand,
	 in the case of large $\eta$, typical in experiments at room temperature, the results may not
	 be reliable due to the fact that in the absence of metallic leads a finite $\eta$ acts as a fictitious reservoir
	 that yields a nonzero dissipation of spin current even in the absence of SOC.
	 A simple approach to rectify the problem is to ignore the effect of finite $\eta$ in the spin pumping term in calculating the Gilbert damping constant.
	 A more accurate approach is to employ  a dephasing mechanism that preserves the conservation laws,
	 which we refer it to as {\it conserving torque correlation} approach discussed in the following subsection.
	
	\subsection{Classical Regime: Conserving Dephasing Mechanism}
 Rather than using the broadening parameter, $\eta$, as a phenomenological parameter, we determine the self energy of the
 Bloch electrons interacting with a dephasing bath associated with phonons, disorder, etc. using a self-consistent Green function approach\cite{Datta2007}. Assuming a momentum-relaxing
 self energy given by,
	\begin{align}\label{eq:4}
	\hat{\Sigma}^{r/a}_{int}(E,t)=\frac{1}{\mathcal{N}}\sum_{k}\hat{\lambda}_{k}\hat{G}^{r/a}_k(E,t)\hat{\lambda}_{k}^{\dagger},
	\end{align}	
	where $\hat{\lambda}_{k}$ is the interaction coupling matrix, the dressed Green function, $\hat{G}^{r/a}_k(E,t)$ ,
	and corresponding self energy, $\hat{\Sigma}^{r/a}_{int}(E,t)$, are calculated self-consistently. This will in turn
	yield a renormalized broadening matrix,
	$\hat{\Gamma}_{int}=\Im(\hat{\Sigma}^r_{int})$, which is the vertex correction modification
	of the infinitesimal initial broadening $\eta_0$.

The nonequilibrium density matrix is calculated from
	\begin{align}\label{eq:5}
	\hat{\rho}^{dis}_{neq}(k;t)=\frac{\hbar}{\pi}\hat{G}_k^r\hat{\Gamma}_{int}\hat{G}_k^a\Big(\frac{\partial \hat{H}_k(t)}{\partial t}+\hat{\mathcal{S}}_t^{aa}\Big)\hat{G}_k^a,
	\end{align}	
	 where the time derivative vertex correction term is
	\begin{align}\label{eq:6}
	\hat{\mathcal{S}}_t^{aa}=\frac{1}{\mathcal{N}}\sum_{k}\hat{\lambda}_{k}\hat{G}^{a}_k\Big(\frac{\partial \hat{H}_k(t)}{\partial t}+\hat{\mathcal{S}}_t^{aa}\Big)\hat{G}^{a}_k\hat{\lambda}_{k}^{\dagger}.
	\end{align}	
	The energy excitation rate for the Bloch electrons then reads,
	\begin{align}\label{eq:8}
	E'_e=\frac{\hbar}{\pi \mathcal{N}}\sum_{k}\Re\Big[\Tr\Big(\Big(\frac{\partial \hat{H}_k(t)}{\partial t}+\hat{\mathcal{S}}_t^{ar}\Big)\hat{\rho}^{dis}_{neq}(k;t)\Big)\Big],
	\end{align}	
	where
	\begin{align}\label{eq:9}
	\hat{\mathcal{S}}_t^{ar}=\frac{1}{\mathcal{N}}\sum_{k}\hat{\lambda}_{k}\hat{G}^{a}_k\Big(\frac{\partial \hat{H}_k(t)}{\partial t}+\hat{\mathcal{S}}_t^{ar}\Big)\hat{G}^{r}_k\hat{\lambda}_{k}^{\dagger}.
	\end{align}	
	The vertex correction Eqs. \eqref{eq:6} and \eqref{eq:9} can be solved
	 either exactly by transforming them into a system of linear equations or by solving them self consistently.
	 Due to the large number of orbitals and atoms per unit cell for the Co/Pt bilayer the latter approach is
	 computationally more efficient.
	In the following numerical calculations we assume $\hat{\lambda}_{k}=\lambda_{int}\hat{1}$ to be a constant independent of $k$ and of orbitals, which can be viewed as the root mean square value of a random on-site potential, $\lambda_{int}=\sqrt{\langle V_{rand}^2\rangle}$, where $\langle...\rangle$ denotes an ensemble averaging, where the self energy in Eq.~\eqref{eq:4} corresponds to the self consistent Born approximation.
	
	\subsection{Gilbert Damping Calculation}
	
	Having determined the energy absorption rate of the Bloch electrons due to the precessing FM, from conservation of energy one can deduce the energy dissipation rate of the FM from, $E'_M=-E'_e$. Using the LLG equation of motion the energy dissipation rate per unit cell of the precessing FM can be obtained from
	\begin{equation}
	E'_M=\frac{1}{2}M_{tot}\hbar\omega\frac{\partial m_z}{\partial t}=-\frac{1}{2}\alpha_{GD} M_{tot}\hbar\omega^2\sin^2(\theta),
	\end{equation}
	where $\vec{m}$ is the unit vector along the magnetization of the FM. The Gilbert damping parameter can then be obtained from
	\begin{equation}
	\alpha_{GD}=\frac{2E'_e}{M_{tot}\hbar\omega^2\sin^2(\theta)}.
	\end{equation}

\section{Computational Scheme}\label{sec:DFTComp}

The spin-polarized density functional theory calculations for the hcp Co(0001)/fcc Pt(111) bilayer were carried out using the Vienna \emph{ab initio} simulation package (VASP) \cite{Kresse96a,Kresse96b}. The pseudopotential and wave functions are treated within the projector-augmented wave (PAW) method \cite{Blochl94,KressePAW}. Structural relaxations were carried using the generalized gradient approximation as parameterized by Perdew {\it{et al.}} \cite{PBE} when the largest atomic force is smaller than 0.01 eV/\AA. The plane wave cutoff energy is 500 eV and a 14 $\times$ 14 $\times$ 1 $k$ point mesh is used in the 2D BZ sampling.
The Pt($m$)/Co($n$) bilayer is modeled employing the slab supercell approach along the [111] consisting of $m$ fcc Pt monolayers (MLs) (ABC stacking) ($m$=1, 2, $\ldots$, 6), $n$ hcp Co MLs (AB stacking) Co ($n$= 6), and a 25 \AA~ thick vacuum region separating the periodic slabs. The in-plane lattice constant of the hexagonal unit cell was set to the experimental value of 2.505~\r{A} for bulk Co.

The Gilbert damping constant was calculated using the tight-binding parameters obtained from VASP-Wannier90 calculations \cite{Mostofi} with a $250\times 250$ $k$-mesh for the bilayer and $250\times 250\times 250$ $k$-mesh for bulk Co. 
The electron Hamiltonian, $\hat{H}_{\vec{k}}$, and exchange splitting, $\hat{\Delta}_{\vec{k}}$, matrices 
in Eq. (\ref{eq:GFeq1}) in the Wannier basis have the form 
\begin{align}
&\hat{H}_{\vec{k}}=\hat{H}_{SOC}+\frac{1}{2}\sum_{\vec{n}}\frac{1}{D_{\vec{n}}}(\hat{H}^{\uparrow\uparrow}_{\vec{n}}+\hat{H}^{\downarrow\downarrow}_{\vec{n}})e^{2i\pi\vec{n}\cdot\vec{k}}\label{eq:param_Hamil}\\
&\hat{\Delta}_{\vec{k}}=\frac{1}{2}\sum_{\vec{n}}\frac{1}{D_{\vec{n}}}(\hat{H}_{\vec{n}}^{\uparrow\uparrow}-\hat{H}^{\downarrow\downarrow}_{\vec{n}})e^{2i\pi\vec{n}\cdot\vec{k}}\label{eq:exch_Hamil},
\end{align}	
where $\hat{H}_{SOC}$ is the SOC Hamiltonian matrix, $\hat{H}^{\uparrow\uparrow}_{\vec{n}}$ and 
$\hat{H}^{\downarrow\downarrow}_{\vec{n}}$ are the spin-majority and spin-minority matrices, 
$\vec{n}=(n_1,n_2,n_3)$ are integers denoting the lattice vectors, 
$D_{\vec{n}}$ is the degeneracy of the Wigner-Seitz grid point, and $k_i\in[0,1]$.

The $\hat{H}^{\uparrow\uparrow}_{\vec{n}}$ and  $\hat{H}^{\downarrow\downarrow}_{\vec{n}}$ are determined from spin-polarized VASP-Wannier90 calculations without SOC.  On the other hand, $\hat{H}_{SOC}$, is determined 
from VASP-Wannier90 non-spin-polarized calculations with SOC as the following.
 
Using the identity,  $Tr[\hat{L}_i\hat{L}_j]=\delta_{ij}\frac{l(l+1)(2l+1)}{3}$, where $\hat{L}_i$ is 
the angular momentum operator of orbital $l$ and $i,j=x,y,z$, the SOC strength of the $I$th atom can be calculated from
\begin{align}
\xi^I_l=\frac{3Tr[\hat{H}^P_{ll,II}\hat{L}_i\hat{\sigma}_i]}{l(l+1)(2l+1)}.
\end{align}	
Here, the superscript $P$ denotes the paramagnetic Hamiltonian, $II$  are the on-site Hamiltonian matrix elements for atom $I$ at $\vec{n}=(0,0,0)$, and $ll$ is the block Hamiltonian matrix corresponding to orbital $l$.  The result is independent of the direction of the angular momentum operator. We find that  $\xi^{Pt}_d=$0.5 eV and $\xi^{Co}_d=$70 meV for the $d$-orbitals of Pt and Co, respectively, that are somewhat smaller than the values considered in the literature\cite{Barati2014} ({\it i.e.} $\xi^{Pt}_d=$0.65 eV, $\xi^{Co}_d=$85 meV).
The SOC Hamiltonian can in turn be written as,
\begin{align}\label{eq:SOC_Hamil}
\langle I,lms|\hat{H}_{SOC}|I',l'm's'\rangle= \frac{1}{2}\delta_{ll'}\delta_{II'}\xi^I_l\sum_i\langle lm|\hat{L}_i|lm'\rangle\hat{\sigma}^i_{ss'}.
\end{align}

\section{Results and Discussion}\label{sec:Results}		%
	\begin{figure}
		\includegraphics[scale=0.4,angle=0]{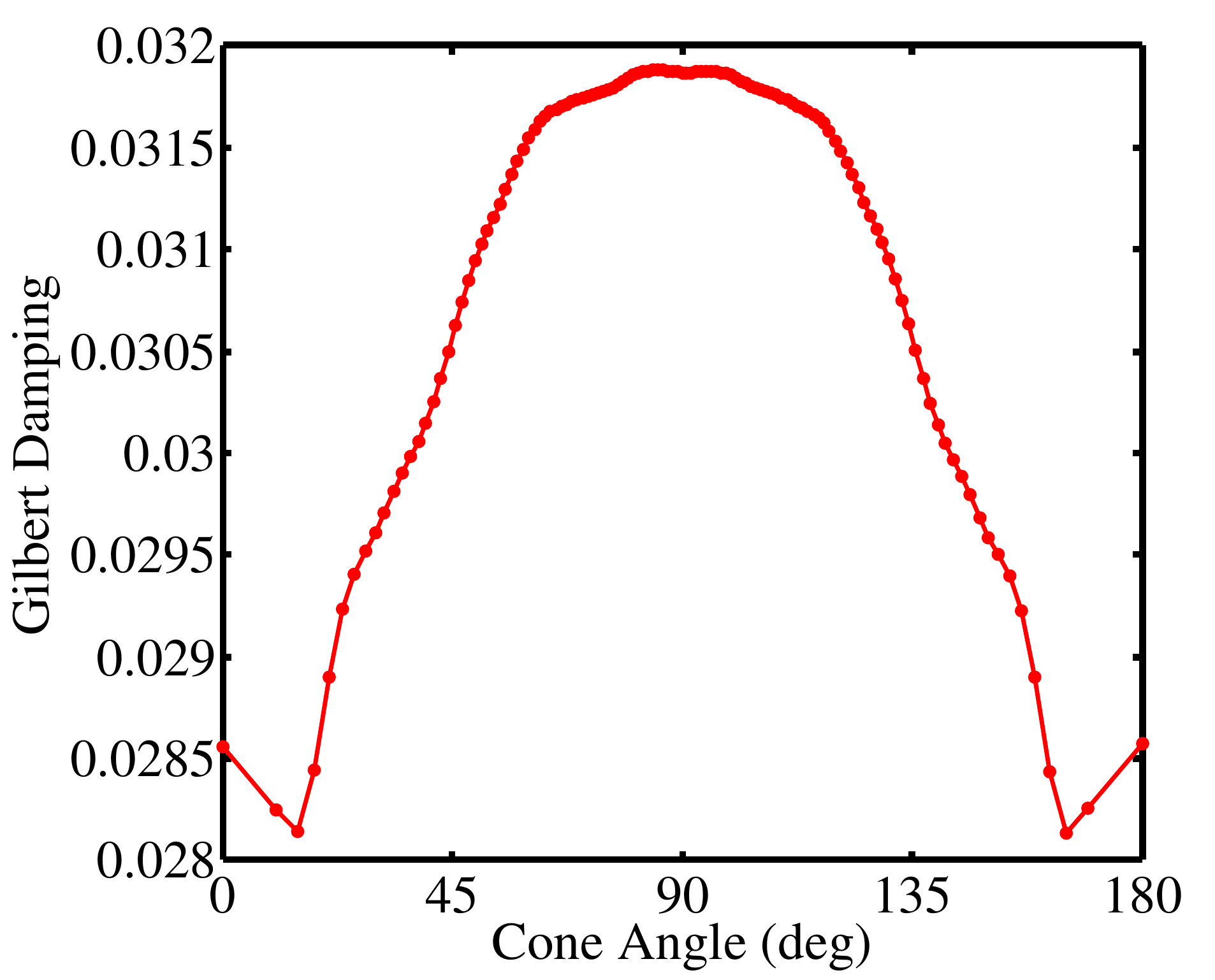}
		\caption{(Color online). Gilbert damping versus precessional cone angle calculated from Eq.~\eqref{eq:GD1} for Pt(1 ML)/Co(6 ML) bilayer system for $S=$60 and $\eta=$1 meV, respectively. }
		\label{fig:fig2}
	\end{figure}
	%	

%%%%%%%%%%%%%%%%%%%%%%%%%%%%%%%%%%%%%%%%%%%%%%%%%%%%  FIG.  2 %%%%%%%%%%%%%%%%%%%%%%%%%%%%%%%%%%
The Gilbert damping constant (calculated from Eq.~\eqref{eq:GD1})versus the precessional cone angle, $\theta_m=\cos^{-1}\Big(\frac{m}{\sqrt{S(S+1)}}\Big)$,
for the Pt(1 ML)/Co(6 ML) bilayer is shown  in Fig.~\ref{fig:fig2} with $\eta=$1 meV and $S=$60. We find that 
the Gilbert damping is relatively independent of the cone angle. The small angular dependence of the Gilbert damping is material dependent and it could increase or decrease upon increasing the cone angle, depending on the material.

%%%%%%%%%%%%%%%%%%%%%%%%%%%%%%%%%%%%%%%%%%%%%%%%%%%%%  FIG. 3  %%%%%%%%%%%%%%%%%%%%%%%%%%%%%%%%%%%%%%%	

\begin{figure}
		\includegraphics[scale=0.4,angle=0]{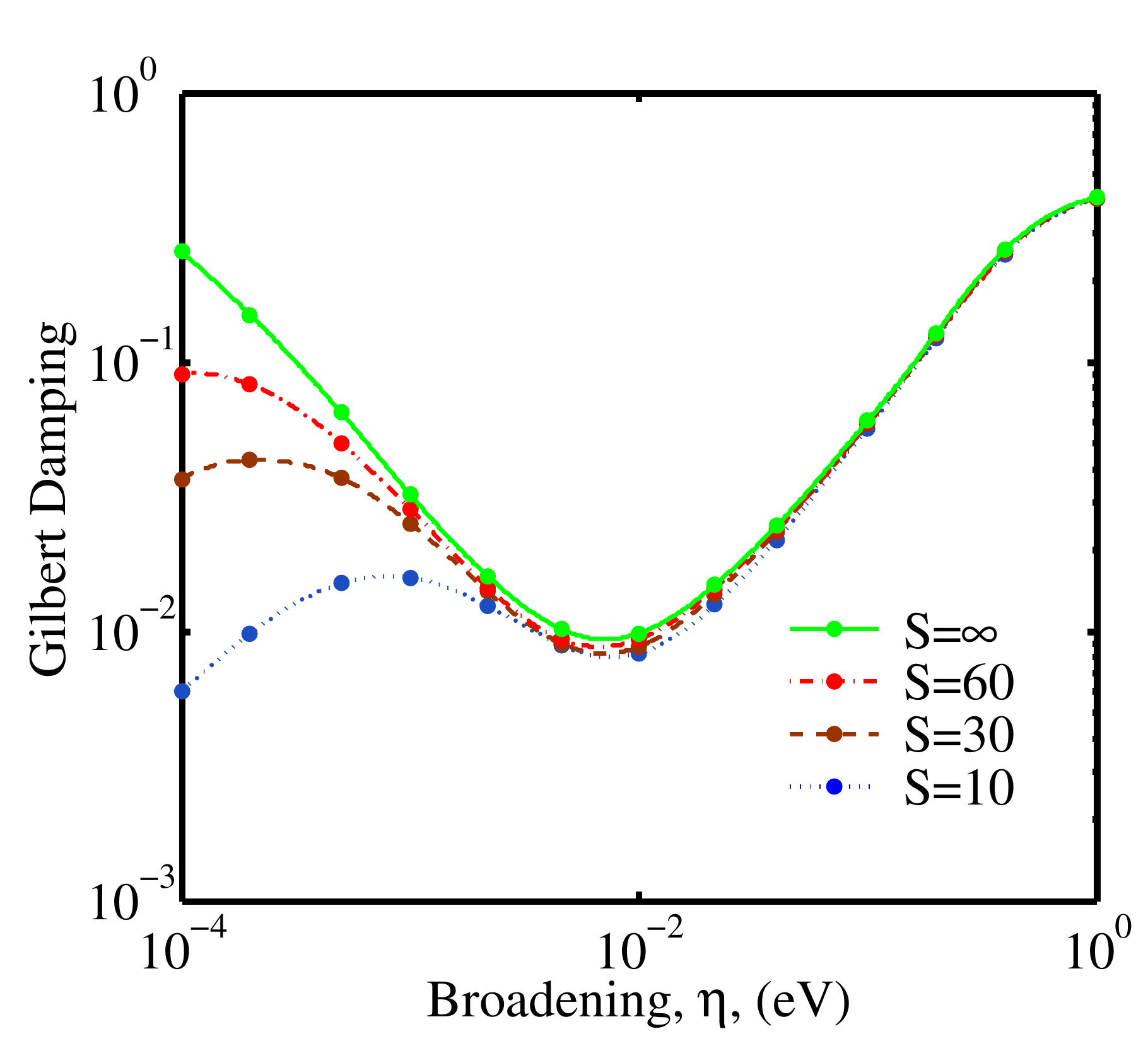}
		\caption{(Color online) Gilbert damping versus broadening parameter for the Pt(1 ML)/Co(6 ML) bilayer 
			for different values of the total spin $S$ of the FM nanocluster. Eqs.~\eqref{eq:GD1} and \eqref{eq:1} were used to 
			calculate the Gilbert damping for finite and infinite $S$, respectively.}
		\label{fig:fig4}
	\end{figure}

In order to see the transition from quantum mechanical to classical dynamical regimes, in Fig.~\ref{fig:fig4} we present the Gilbert damping constant of the Pt(1 ML)/Co(6 ML) bilayer versus the broadening parameter, $\eta$, for different values of the total spin $S$ of the FM. For the case of finite $S$ we used Eq.~\eqref{eq:GD1} while 
 for $S=\infty$ we used the TC expression  Eq.~\eqref{eq:1}. We find that for finite $S$ the Gilbert damping value exhibits a peak in the 
 small $\eta$ regime (clean system) where the peak value increases linearly with $S$ and shifts to smaller broadening value with increasing $S$.
The underlying origin of the $\alpha_{GD}$($\eta$) behavior with $S$ in coherent regime can be understood in terms of the coherent transport of quasi-particles along the auxiliary direction $m$ in Fig.~\ref{fig:fig1}, where the auxiliary current flow (damping rate) depends linearly on the chemical potential difference between the first ($m=+S$) and last layers ($m=-S$) which is simply 2$S$. This suggests that in the limit of infinite $S$ and ballistic regime $\eta\rightarrow 0$ the intrinsic Gilbert damping diverges. It was shown that the problem of infinite Gilbert damping in the ballistic regime can be removed by taking into account the collective excitations.~\cite{Costa2015,Edwards2016}
	%	
	
\begin{figure}
		\includegraphics[scale=0.4,angle=0]{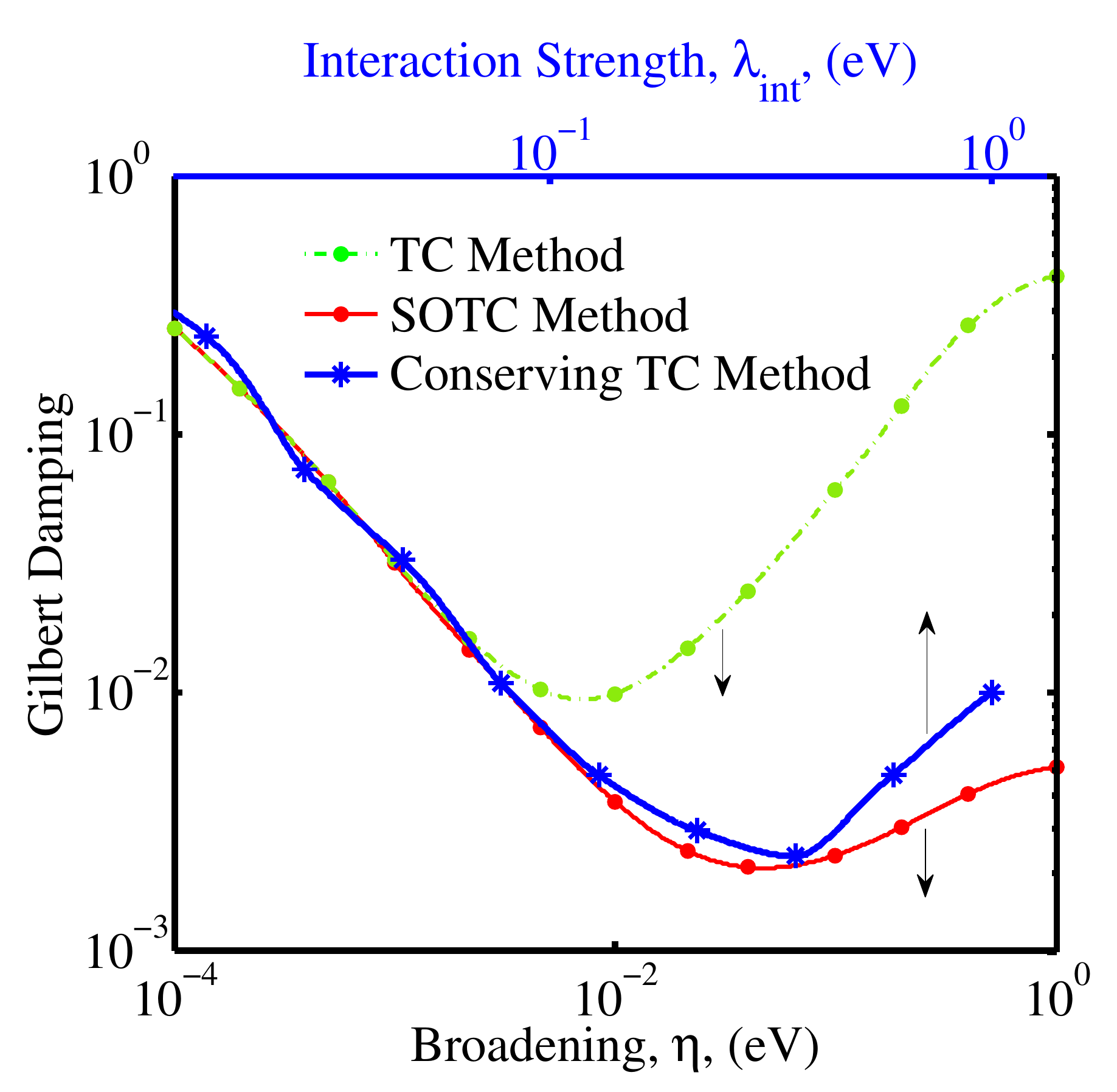}
		\caption{(Color online). Gilbert damping of Pt(1 ML)/Co(6 ML) bilayer versus the broadening parameter $\eta$
			(lower abscissa) and interaction strength, $\lambda_{int}$, (upper abscissa), using the torque correlation (TC), spin-orbital torque correlation (SOTC), and conserving TC expressions given by Eqs.~\eqref{eq:1}, \eqref{eq:3} and \eqref{eq:8}, respectively.}
		\label{fig:fig3}
	\end{figure}

	%%%%%%%%%%%%%%%%%%%%%%%%%%%%%%%%%%%%  DIFFERENT METHODS FOR GILBERT DAMPING
	As we discussed in Sec.~\ref{sec:2B}, the relaxation time approximation is valid only in the small $\eta$ limit and it violates the conservation law when $\eta$ is large. In order to quantify the validity of the relaxation time approximation, in Fig.~\ref{fig:fig3} we display 
 the Gilbert damping versus broadening, $\eta,$ or interaction parameter, $\lambda_{int}$, for the Pt(1 ML)/Co(6 ML) bilayer 
	with $S=\infty$, using {\it (i)} torque correlation(TC) expression (Eq.~\eqref{eq:1}); {\it (ii)} the spin-orbital torque correlation (SOTC) expression (second term in Eq.~\eqref{eq:3};) and {\it (iii)} the conserving TC expression (Eq.~\eqref{eq:8}). The upper horizontal-axis refers to the interaction strength $\lambda_{int}$ of the conserving TC method and the lower one refers to the broadening parameter, $\eta$.
	The calculations show that for $\eta>$20 meV the TC results deviate substantially from those of the conserving TC method. Ignoring the spin pumping contribution to the Gilbert damping in Eq.~\eqref{eq:3} and considering only the SOTC component increases the range of the validity of the relaxation time approximation. 
	Therefore, the overestimation of the Gilbert damping using the TC method can be attributed to the disappearance of electrons (pumped spin current) in the presence of the finite non-Hermitian term, $i\eta\hat{1}$, in the Hamiltonian.
	%	
	
\begin{figure}
		\includegraphics[scale=0.4,angle=0]{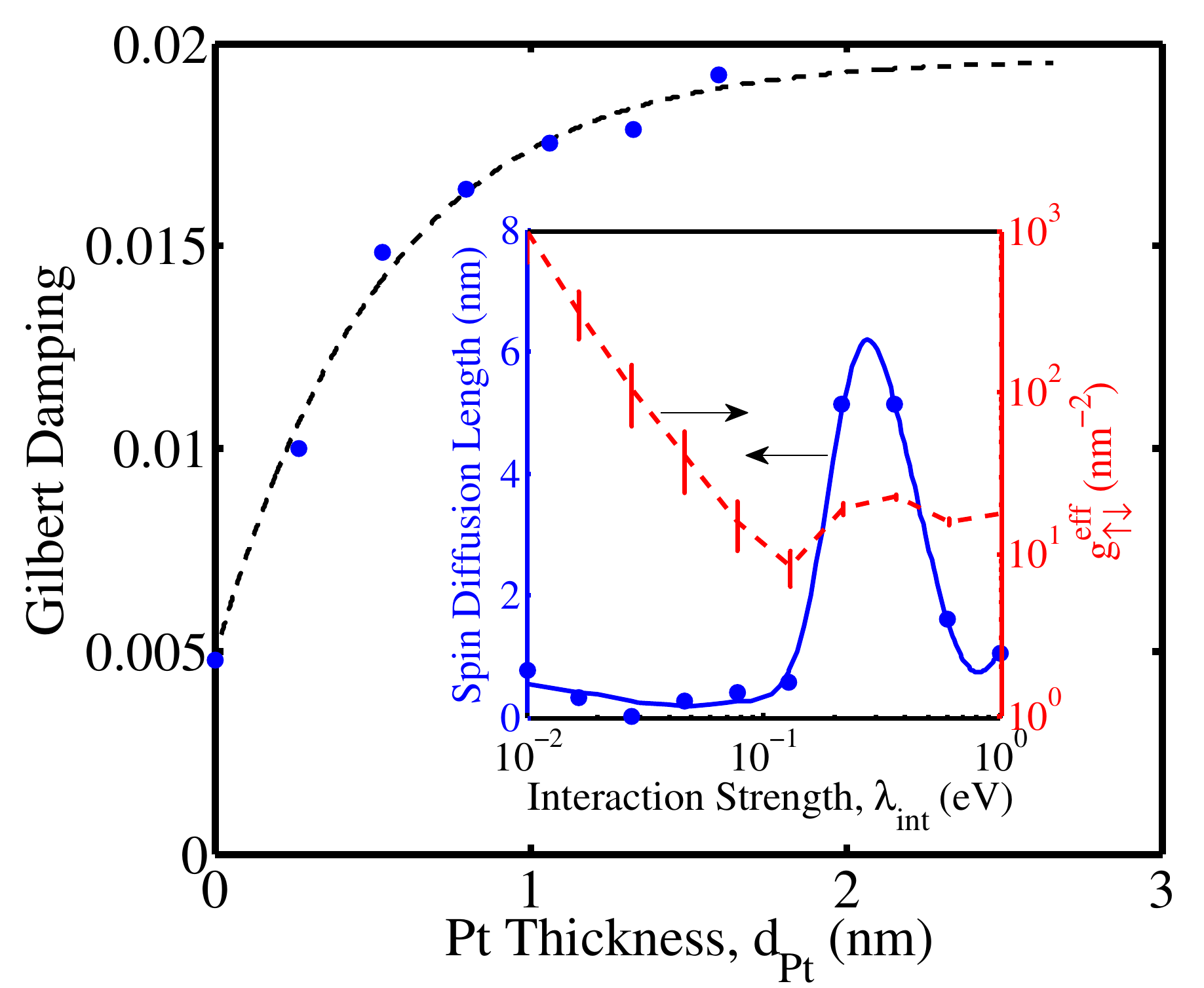}
		\caption{(Color online). {\it Ab initio} values (circles) of Gilbert damping versus Pt thickness for Pt($m$ ML)/Co(6 ML) bilayer where $m$ ranges between 0 and 6 and $\lambda_{\textrm{int}}=1 eV$. The dashed curve is the fit of the Gilbert damping values to Eq. (\ref{eq:Sp_diff}). 
			Inset: spin diffusion length (left ordinate) and effective spin mixing conductance, $g^{\textrm{eff}}_{\uparrow\downarrow}$, (right ordinate) versus interaction strength. The errorbar for $g^{\textrm{eff}}_{\uparrow\downarrow}$ is equal to the root mean square deviation of the damping data from the fitted curve.}
		\label{fig:fig5}
	\end{figure}

	%%%%%%%%%%%%%%%%%%%%%%%%%%%%%    SPIN DIFFUSION LENGTH AND SPIN MIXING CONDUCTANCE %%%%%%%%%%%%%%%%%%%%%
	We have used the conserving TC approach to calculate the effect of $\lambda_{int}$ on  
	the Gilbert damping as a function of the Pt layer thickness for the Pt($m$)/Co(6 ML) bilayer. 
	As an example, we display in Fig.~\ref{fig:fig5} the results of Gilbert damping versus Pt thickness for $\lambda_{int}=1 eV$ which yields a Gilbert damping value of ~0.005 for bulk Co ($m$ = 0 ML) and is in the range of  ~0.005\cite{Bhagat1974,Kato2012} to 0.011\cite{Rojas2014,Kim2016,Rana2011} reported experimentally.
	Note that this large $\lambda_{int}$ value describes the Gilbert damping in the resistivity-like regime which might not be appropriate to experiment, where the bulk Gilbert damping decreases with temperature, suggesting that it is in the conductivity regime.\cite{Isasa2015}
	
	For a given $\lambda_{int}$ we fitted the 
	{\it ab initio} calculated Gilbert damping versus Pt thickness to the spin diffusion model,\cite{Foros2005,Shaw2012,Ghosh2012}
	%%%%%%%%%%%%%%%%%%%%%%%%%%%%%%%%%%%%%%%%%%%%%%%%%%
	\begin{align}\label{eq:Sp_diff}
	\alpha_{Pt/Co}=\alpha_{Co}+\frac{g^{\textrm{eff}}_{\uparrow\downarrow}V_{Co}}{2\pi M_{Co}d_{Co}}(1-e^{-2d_{Pt}/L^{sf}_{Pt}}).
	\end{align}	
	Here, $g^{\textrm{eff}}_{\uparrow\downarrow}$ is the effective spin mixing conductance, $d_{Co}$ ($d_{Pt}$ ) is the thickness of Co (Pt), $V_{Co}=10.5~\textrm{\r{A}}^3$ ($M_{Co}=1.6 \mu_B$) 
	 is the volume (magnetic moment) per atom in bulk Co, and $L^{sf}_{Pt}$ is the spin diffusion length of Pt.
	The inset of Fig.~\ref{fig:fig5} shows the variation of the effective spin mixing conductance 
	and spin diffusion length with the interaction strength $\lambda_{int}$.
	In the diffusive regime $\lambda_{int}>0.2 eV$, $L^{sf}_{Pt}$ ranges between 1 to 6 nm in agreement with experiment findings which are between 0.5 and 10 nm\cite{Rojas2014,Boone2013}. 
	Moreover, the effective spin mixing conductance is relatively independent of $\lambda_{int}$  oscillating around 20 nm$^{-2}$, which is approximately half of the experimental value of $\approx$ ~35 - 40 nm$^{-2}$.\cite{Rojas2014,Azzawi2016}
	On the other hand, in the ballistic regime ($\lambda_{int}<$0.2 eV), although the errorbar in fitting to the diffusion model is relatively large, the value of $L^{sf}_{Pt}\approx$ 0.5 nm is in agreement with Ref.\cite{Barati2014} and experimental observation \cite{Boone2013}.

	%	
%%%%%%%%%%%%%%%%%%%%%%%%%%%%%%%%%%%%%%%%%%%%%%%%%%%%%%%%%%%%%%%%%%%%%%%%%%%%%%%%%%%%%%%%%%%%%%%%%%%%%%
	\section{Concluding remarks}\label{sec:conclusions}
	We have developed an {\it ab initio}-based electronic structure framework to study the magnetization dynamics of a nano-FM  where its magnetization is treated quantum mechanically. The formalism was applied to investigate the intrinsic Gilbert damping of a Co/Pt bilayer as a function of energy broadening. We showed that in the limit of small $S$ and ballistic regime the FM damping is governed by coherent dynamics, where the Gilbert damping is proportional to $S$. In order to study the
	effect of disorder on the Gilbert damping we used a relaxation scheme within the self-consistent Born approximation. The {\it ab initio} calculated Gilbert damping as a function of Pt thickness were fitted to the spin diffusion model for a wide range of disorder strength. In the limit of large disorder strength the calculated spin diffusion length and effective spin mixing conductance are in relative agreement with experimental observations.	
	
	\begin{acknowledgments}
		
	The work is supported by NSF ERC-Translational Applications of Nanoscale Multiferroic Systems (TANMS)- Grant No. 1160504 and by NSF-Partnership in Research and Education in Materials (PREM) Grant No. DMR-1205734.
	\end{acknowledgments}

	%BibTeX
	%Windows:
	%\bibliographystyle{D:/PHYSICS/TEX/BIBTEX/prsty}
	%\bibliography{D:/PHYSICS/TEX/BIBTEX/qttg}
	
	%Linux:
	%\bibliographystyle{apsrev}
	%\bibliography{$HOME/TEX/BIBTEX/qttg}
	
\end{document}